\documentclass[]{krakproc}
\usepackage{graphicx}
\usepackage{wrapfig}
\newcommand{\msolar}{{\rm M_{\odot}}}

\newcommand{\hi}{{\rm H{\sc i}\,}}
\newcommand{\ovi}{{\rm O{\sc vi}\,}}

\newcommand{\novi}{{\rm N(O{\sc vi})\,}}
\newcommand{\nhi}{{\rm N(H{\sc i})\,}}

\begin{document}

\title{Global MHD Modelling of the ISM - From large towards small scale turbulence}
\author{Miguel A. de Avillez$^{1,2}$ and Dieter Breitschwerdt$^{1}$}
\institute{$^{1}$Institut f\"ur Astronomie, Universit\"at Wien, T\"urkenschanzstra\ss e 17, A-1080 Wien, Austria\\
$^{2}$Department of Mathematics, University of \'Evora, R. Rom\~ao Ramalho 59, 7000 \'Evora, Portugal}
\markboth{M. A. de Avillez \& D. Breitschwerdt}{Global MHD Modelling of the ISM}

\maketitle

\begin{abstract}

Dealing numerically with the turbulent nature and non-linearity of the
physical processes involved in the ISM requires the use of
sophisticated numerical schemes coupled to HD and MHD mathematical
models. SNe are the main drivers of the interstellar turbulence by
transferring kinetic energy into the system. This energy is dissipated
by shocks (which is more efficient) and by molecular viscosity. We
carried out adaptive mesh refinement simulations (with a finest
resolution of 0.625 pc) of the turbulent ISM embedded in a magnetic
field with mean field components of 2 and 3 $\mu$G. The time scale of
our run was 400 Myr, sufficiently long to avoid memory effects of the
initial setup, and to allow for a global dynamical equilibrium to be
reached in case of a constant energy input rate. It is found that the
longitudinal and transverse turbulent length scales have a time
averaged (over a period of 50 Myr) ratio of 0.52-0.6, almost similar
to the one expected for isotropic homogeneous turbulence. The mean
characteristic size of the larger eddies is found to be $\sim 75$ pc
in both runs. In order to check the simulations against observations,
we monitored the \ovi and \hi column densities within a superbubble
created by the explosions of 19 SNe having masses and velocities of
the stars that exploded in vicinity of the Sun generating the Local
Bubble. The model reproduces the FUSE absorption measurements towards
25 white dwarfs of the \ovi column density as function of distance and
of \nhi. In particular for lines of sight with lengths smaller than
120 pc it is found that there is no correlation between \novi and
\nhi.
\end{abstract}

\section{Introduction}
The Reynolds number of the ISM is of the order of $10^{6}$ or even
higher. Thus, the ISM is a turbulent medium that is regularly stirred
by shock waves generated in supernova explosions. At least $10\%$ of
energy from SN blast waves is injected into the interstellar gas
(Thornton et al. 1998; however, for higher efficiencies see Avillez \&
Breitschwerdt 2005c). The energy injected into turbulence at the
largest scale is transferred to smaller and smaller eddies until it
reaches the Kolmogorov inner scale of the order of $10^{-3}$ pc, or
even lower.

The interstellar turbulence, which is mainly due to shear flows of
large scale streams and its associated increase of vorticity, decays
without constant energy supply. Attempts have been made to
understand the time scales of the turbulence decay within molecular
clouds and it was found that the presence of magnetic fields would
only delay this decay (Mac Low et al. 1998). These studies use a
computational domain having the size of the turbulence outer scale
(the scale at which energy is injected). However, although molecular
clouds can have sizes of a few tens of parsecs, they are embedded in
a large scale interstellar medium (ISM) that is swept up by shock
waves from SNe, driving large scale streams that eventually generate
shock compressed layers leading to the formation of molecular
clouds. In addition, the occupation fraction of the hot gas in the
Galactic disk is low (some 20\%; Avillez \& Breitschwerdt 2004,
2005a) because there are kpc-scale flows escaping into the thick disk
or into the halo through chimneys. These flows enter the
disk-halo-disk cycle becoming later a source of turbulence in the
disk as the \hi clouds, generated in the fountain, strike the disk
and enhance the vorticity.

Modelling the ISM in a self-consistent way requires an approach, that
takes into account the relevant turbulent scales, which cover several
orders of magnitude as well as the disk-halo-circulation. This is a
difficult task that requires the use of sophisticated numerical codes,
adequate computing power, and precision input data by
observations. Only since very recently, by the rapid evolution of telescope
and detector technology, as well as the availability of large numbers
of parallel processors, we are in the fortunate situation to follow in
detail the evolution of the ISM on the global scale taking into
account the disk-halo-disk circulation in three dimensions.

The plan of the present paper is as follows. In Section 2 a brief
overview of the developments carried out in the 3D supernova-driven
ISM model of Avillez (2000) (in order to provide a self-consistent
picture of the ISM) is presented. Some results on turbulence scales
and tests to the present model are discussed in Sections 3 and 4,
respectively. In Section 5 a few final remarks are presented.

\section{Modelling of a Supernova-driven ISM}

The supernova-driven model coupled to a 3D
magneto-hydrodynamical block mesh refinement scheme used to study the
properties of the ISM in general (e.g., Avillez \& Berry 2001; Avillez
\& Mac Low 2001, 2002; Avillez \& Breitschwerdt 2004, 2005a;
Breitschwerdt \& Avillez, these proceedings) and of the Local Bubble
(for a review see Breitschwerdt 2001), in particular (Avillez \&
Breitschwerdt 2005b), is based on the Ph.D. thesis of Avillez
(1998). 

The present model evolved from a primitive 3D single grid code, that
solved the Euler equations and included local heating by SNe types
Ib+c and II and radiative cooling assuming collisional ionization
equilibrium in an optically thin ISM. Soon it was realized that the
lack of resolution affected the simulations, as small scale phenomena
were inhibited, which in turn affected the cooling.  Furthermore, the
turbulence inner and outer scales differ by several orders of
magnitude.  Dealing simultaneously with such large differences in
scales requires the use of high resolution grids.  However, increasing
the grid resolution would make the simulations impossible, unless grid
refinement on the fly (in regions of the flow where pressure and/or
density gradients occur) in a parallel fashion is used. The present
code includes magnetic fields using the finite difference scheme of
Dai \& Woodward (1998) and a block refinement algorithm based on
Berger \& Collela (1989) and on Balsara (2001), assuring conservation
of $\nabla\cdot \vec{B}=0$ during grid refinement. With this
code 3D HD and MHD simulations of the ISM are capable of reaching
resolutions as high as 0.625 pc (a couple of test runs were even
carried out with a resolution of 0.3125 pc), while the coarse grid
resolution is 10 pc.

At the time of writing the present paper, the code already
incorporates self-gravity and cosmic rays. The latter setup is quite
different from that described in Hanasz et al. (2003) and Ryu et al.
(2003), as the wave field interaction with cosmic rays is taken into
account, allowing for resonant wave generation by the cosmic ray
streaming instability. Further developments are underway, such as the
inclusion of non-equilibrium ionization, thermal conduction and
differential rotation (see Figure~\ref{hist} for a sketch of the time
evolution of the model).

\begin{figure}[th]
\centering
\includegraphics[width=1.\hsize,angle=0]{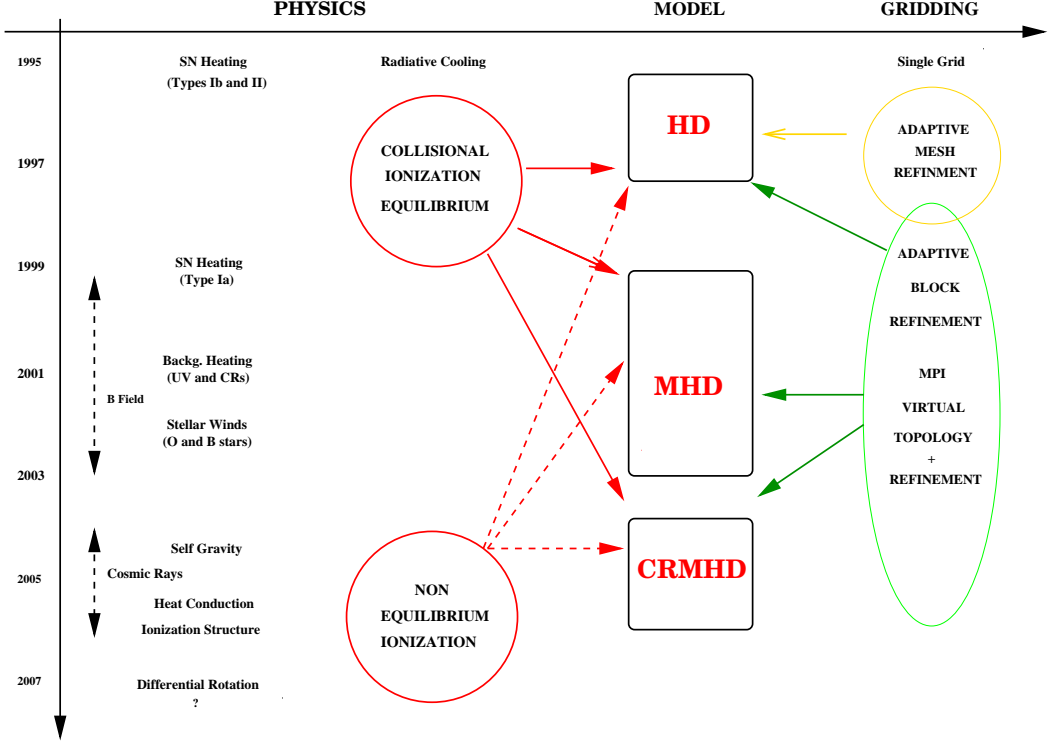}
\caption{Structure and time schedule of the supernova-driven ISM model.
\label{hist}
}
\end{figure}

\section{Characterizing Turbulence Scales in the ISM}

\begin{figure}[thbp]
\centering
\includegraphics[width=0.45\hsize,angle=0]{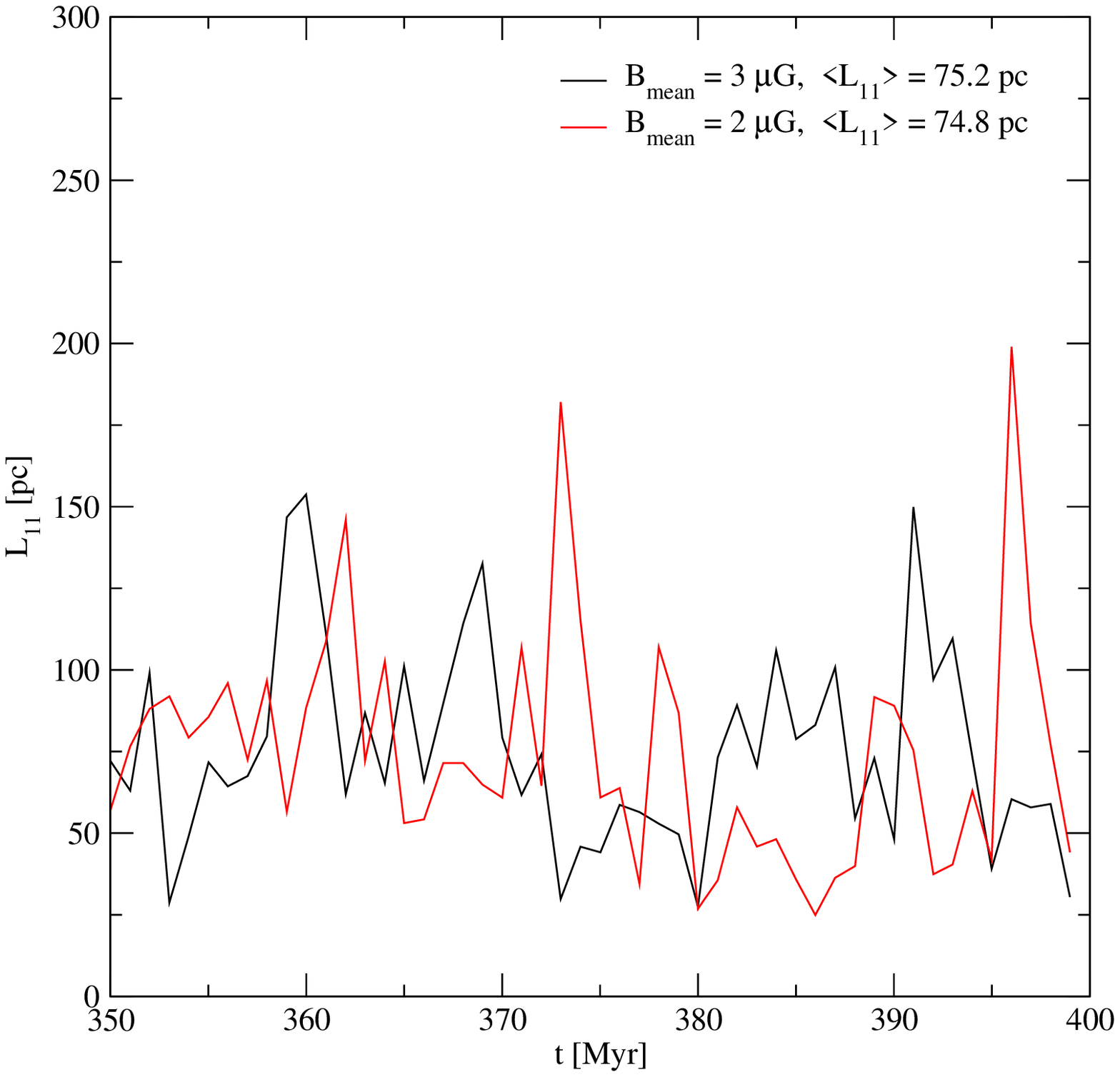}
\includegraphics[width=0.45\hsize,angle=0]{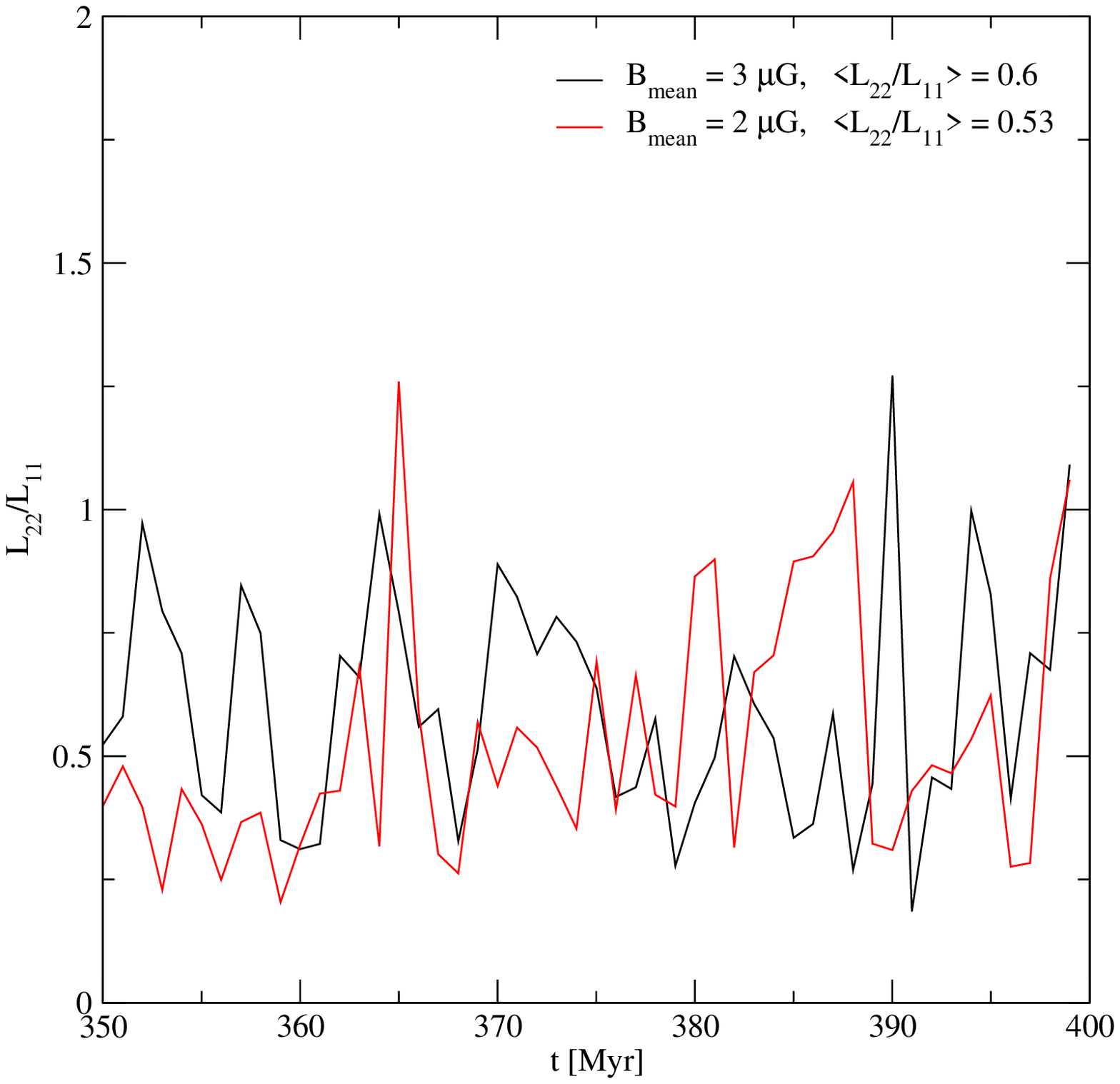}
\caption{History of the characteristic size (given by $L_{11}$) of
the larger eddies (left panel) and of the ratio $L_{22}/L_{11}$
(right panel) for the runs with mean field strengths of 2 (red) and
3 (black) $\mu$G. }
\label{l11}
\end{figure}

The model can be used to characterize turbulence in the ISM and
determine the integral scales at which the gas correlates both in time
and space. The outer scale of the turbulent flow in the ISM is related
to the scale at which the energy in blast waves is transferred to the
interstellar gas. Such a scale can be determined by using the
so-called two-point correlation function $R_{i,j}(\vec{r},t)=\langle
u_{i}(\vec{x}+\vec{r},t)u_{j}(\vec{x},t)\rangle$, which in the case of
isotropic turbulence can be written in terms of the scalar functions
$f(r,t)$ and $g(r,t)$ as
\begin{equation}
R_{ij}(\vec{r},t)=\sigma^{2}(u)\left(g(r,t)\delta_{ij}+\left[f(r,t)-g(r,t)\right]\frac{r_{i}r_{j}}{r^{2}}\right),
\end{equation}
where $\sigma^{2}(u)$ and $r_{i}$ are the variance of the velocity
and the $i-$component of the displacement vector $\vec{r}$. If
$\vec{r}$ is chosen to be oriented along the mean magnetic field
($x-$axis), $\vec{r}=r \vec{e}_{x}$, then
\begin{equation}
R_{11}/\sigma^{2}(u)=f(r,t), \quad R_{33}=R_{22}/\sigma^{2}(u)=g(r,t),~\mbox{and}~ R_{ij}=0~\mbox{for}~i\neq j.
\end{equation}
$R_{11}$ and $R_{22}=R_{33}$ are the longitudinal and transverse
autocorrelation functions. The characteristic size of the larger
eddies in the flow is given by the area under the curve of the
autocorrelation function
\begin{equation}
L_{11}=\int_{0}^{+\infty}f(r,t) dr.
\end{equation}
The left panel of Figure~\ref{l11} shows the history of $L_{11}$ in
the last 50 Myr of evolution of the magnetized ISM with mean field
strengths of 2 (red) and 3 $\mu$G (black). The average integral
length scales in both cases is 75 pc. There is a large scatter of
$L_{11}$ around its mean as a result of oscillations in the
momentary star formation rate, which is determined locally by
density and temperature thresholds (note that in these simulations
the mean SN rate is similar to the Galactic value), and on the
formation of superbubbles, being responsible for the peaks observed
in the two plots. The transverse integral length scale, $L_{22}$, given by
\begin{equation}
L_{22}=\int_{0}^{+\infty}g(r,t) dr = 0.5 L_{11}
\label{eql22}
\end{equation}
in isotropic turbulence.
 However, in the present MHD simulations
$0.2<L_{22}/L_{11}<1.3$ (right panel of Figure~\ref{l11}) for the two
mean field strengths. In spite of the large scatter the time average
of the $L_{22}/L_{11}$ over the 50 Myr period is 0.53 and 0.6 for
$B_{mean}=2$ and $3~\mu$G, respectively, a value similar to the one
predicted by (\ref{eql22}). A similar behaviour is seen in the
correlation functions calculated for the HD case, that is, there is a
fluctuation of the ratio $L_{22}/L_{11}$ (Avillez \& Breitschwerdt
2005c). In a statistical sense turbulence seems to be homogeneous and
isotropic, but the history of the $L_{22}/L_{11}$ ratio indicates that
this does not occur in general, and therefore, turbulence in the
magnetized (and unmagnetized) ISM does not perfectly satisfy this.

It should be kept in mind that the relations (3) and (4) assume an
isotropic and homogeneous (in the local sense) turbulent medium
where energy is injected at the outer scales and transferred to the
smallest scales, without any dissipation, until it reaches the
Kolmogorov inner scale given by $R_{e}^{-3/4}l_{o}$ ($l_{o}$ is the
injection scale), where it is dissipated by the molecular viscosity.
Dissipation is a passive process as it proceeds at a rate determined
by the inviscid inertial behaviour of the large eddies. Such an
energy cascading corresponds to a divergence-free behaviour of the
flow in the inertial range.

However, the ISM is a compressible medium swept up by shocks, which
are in general more efficient in dissipating energy than molecular
viscosity. Simulations of decaying sonic (Porter et al. 1998) and
forced supersonic (Boldyrev et al. 2002) turbulence
indicate that the ratio $\gamma=\langle u_{c}^{2} \rangle/\langle
u_{s}^{2} \rangle$ that relates the compressional $u_{c}$ to the
selenoidal $u_{s}$ component of the velocity field in the inertial
range is smaller than 0.2. Thus, a small amount of energy is
dissipated through shocks in the inertial range and the energy decay
follows a quasi-Kolmogorov spectrum. The shock structures start to
play an important r\^ole in energy transfer and dissipation near the
dissipative range (Boldyrev 2002).

\begin{figure}[thbp]
\centering
\vspace*{3.2cm}
Jpeg image 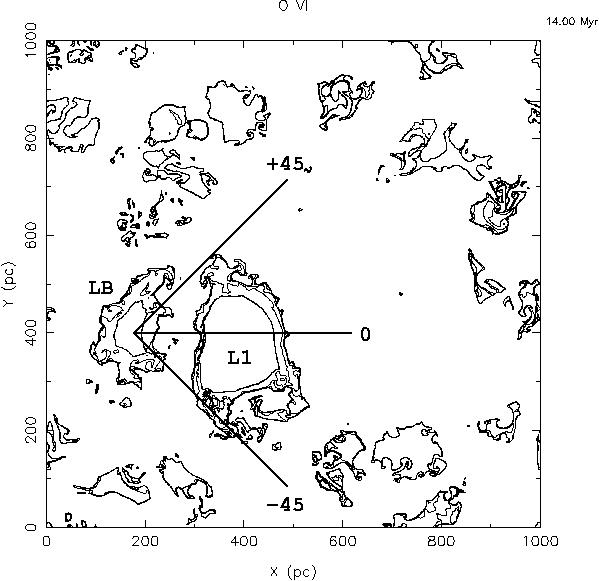
 \vspace*{3.2cm}
\caption{ \ovi contour map of a 3D Local Bubble simulation 14 Myr
after the
 first explosion in the cluster. The LB is centred at (175, 400) pc
 and Loop I at (375, 400) pc.
\label{lb}
}
\end{figure}
\section{Testing the Model}
How much can the present SN driven ISM model setup be trusted? The
answer to this question relies on the tests made with the numerical
schemes and the physics approaches used in the ISM setup. The
standard numerical tests carried out on the HD and MHD schemes
showed that the code produces solutions, under different
gas-dynamical conditions, that differ from the expected analytical
and/or experimental solutions by at most $1\%$. Lately it has been
claimed that a fundamental test of ISM models is the determination
of the amount of \ovi column densities within a superbubble, similar
to the Local Bubble (LB) (see Cox 2004; Breitschwerdt \& Cox 2004).
Observations along lines of sight near the disk within the Local
Bubble and extending further away show that the \ovi column density
(\novi) has a mean value of $8\times 10^{12}$ cm$^{-2}$, with no
values larger than $1.7\times 10^{13}$ cm$^{-2}$ according to FUSE
observations (Oegerle et al. 2005). Therefore, in absorption, a
fundamental test is the setup of the LB and the prediction of the
\ovi and \hi column densities and the \emph{direct }comparison with
observed column densities.

Thus, we ran simulations following the evolution of the Local Bubble
and Loop I as a result of clustered SN activity in an ISM disturbed
by SN explosions at the Galactic rate. For this we took data cubes
of previous runs with a finest adaptive mesh refinement resolution
of 1.25 pc. We then picked up a site with enough mass to form the 81
stars, with masses, $M_*$, between 7 and 31 $\msolar$, that compose
the Sco Cen cluster; 39 massive stars with $14\leq M_* \leq 31 \,
\msolar$ have already gone off, generating the Loop I cavity.
Presently the Sco Cen cluster (here located at $(375,400)$ pc) has
42 stars to explode within the next 13 Myrs). We followed the
trajectory of the moving subgroup B1 of Pleiades, whose SNe in the
LB went off along a path crossing the solar neighbourhood
(Figure~\ref{lb}). Periodic boundary conditions are applied along
the four vertical boundary faces, while outflow boundary conditions
are imposed at the top ($z=10$ kpc) and bottom ($z=-10$ kpc)
boundaries. The simulation time of this run was 30 Myr, sufficiently
long to cover the total evolution of both bubbles.

The locally enhanced SN rates produce coherent LB and Loop I
structures (due to ongoing star formation) within a highly disturbed
background medium. The successive explosions heat and pressurize the
LB, which at first looks smooth, but develops internal temperature and
density structure at later stages. After 14 Myr the LB cavity, bounded
by an outer shell, which will start to fragment due to Rayleigh-Taylor
instabilities in $\sim 3$ Myr from now, fills a volume roughly
corresponding to the present day size (Fig.~\ref{lb}).

\begin{figure}[thbp]
\centering
\includegraphics[width=0.5\hsize,angle=0]{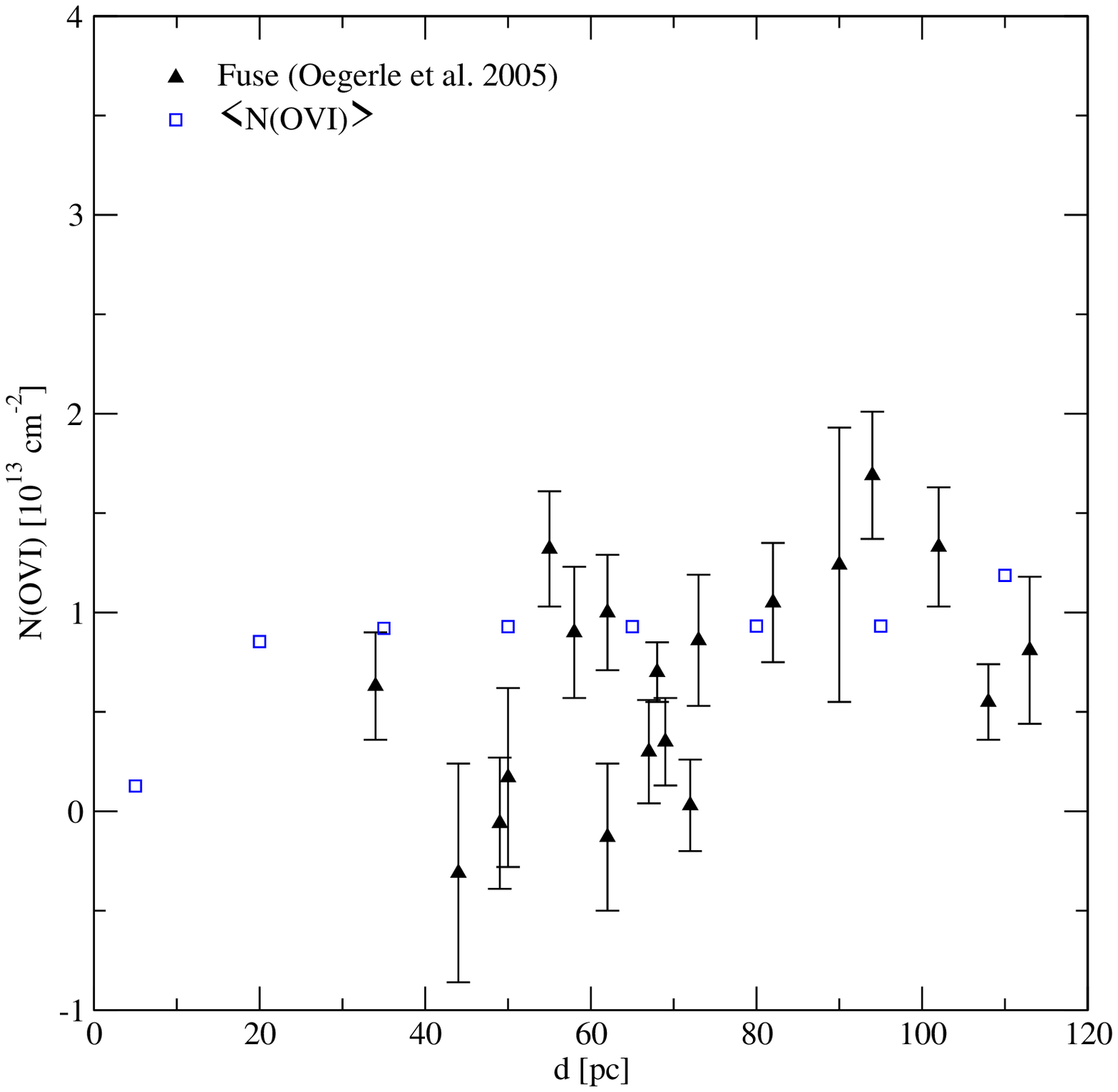}\includegraphics[width=0.5\hsize,angle=0]{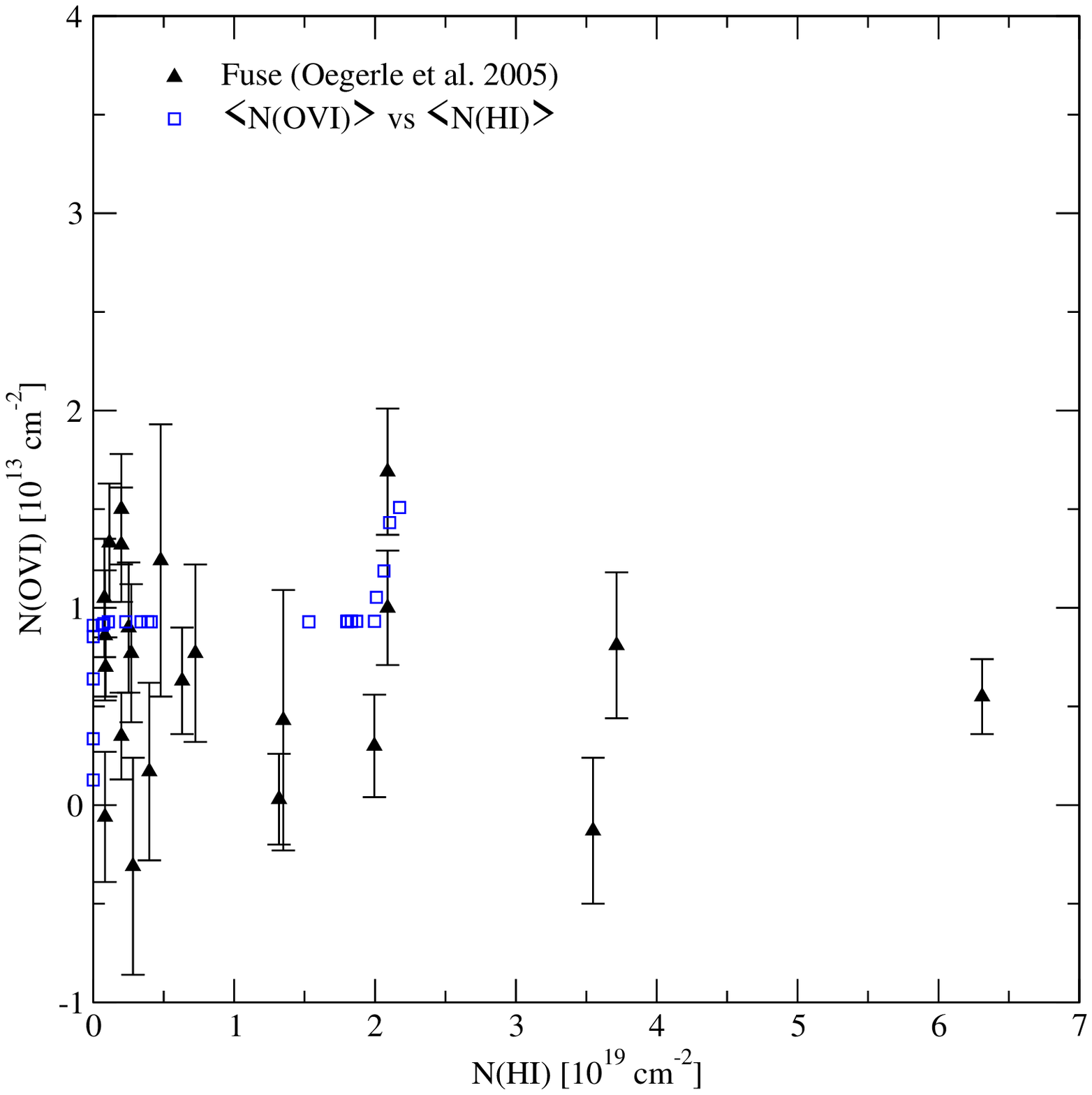}
\caption{Variation of the {\em calculated} averaged \ovi column
density with the distance (left panel) and with the {\em calculated}
averaged \hi column density (right panel) within the Local Bubble
for lines of sight with length $< 120$ pc, and compared to FUSE
observations.} \label{ovi}
\end{figure}

The average \ovi and \nhi column densities, i.e., $\langle
\mbox{\novi} \rangle$ and $\langle\mbox{\nhi}\rangle$, respectively,
are calculated by averaging the column densities measured along 91
lines of sight (LOS) extending from the Sun and crossing Loop I from
an angle of $-45^{\circ}$ to $+45^{\circ}$ (as shown in
Figure~\ref{lb}) with angle increments of $1^{\circ}$. The left
panel of Figure~\ref{ovi} shows $\langle\mbox{\novi} \rangle$ within
the Local Bubble (i.e., for a LOS length $l_{LOS}< 120$ pc) as
function of distance (left panel) and of $\langle\mbox{\nhi}\rangle$
(right panel) at time 14.5 Myr of evolution after the first
explosion and 0.5 Myr after the last explosion. Overlayed to both
panels is the corresponding \ovi FUSE data taken from Oegerle et al.
(2005). The clustered SN driven LB cavity simulated data shows that
the averaged \ovi column density falls within the observed data and
there is excellent agreement with respect to the observed distances.
A similar result is found for the minimum and maximum column
densities (see Breitschwerdt \& Avillez 2005).

\begin{wrapfigure}[21]{l}[0pt]{7.5cm}
\centering
\centerline{\includegraphics[width=0.9\hsize,angle=0]{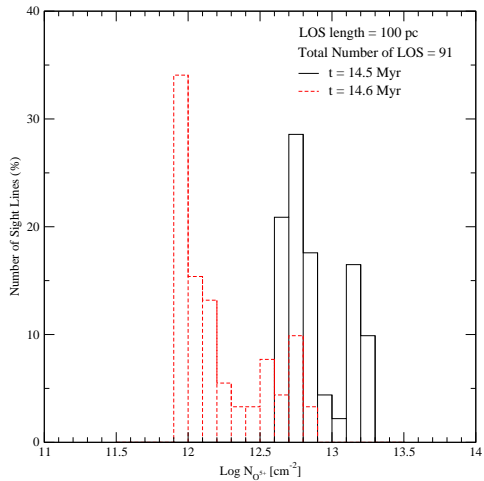}}
\begin{minipage}[l]{7cm}
\caption{ Histogram of the percentage of LOS with various ranges of
observed $\mbox{N(\ion{O}{vi})}$ within the LB at $t=14.5$ and
$14.6$ Myr.}
\end{minipage}
\label{hist}
\end{wrapfigure}
Histograms (Figure~\ref{hist}) of column densities
obtained in the 91 LOS for $t=14.5$ show that 67\% of the lines have
column densities smaller than $10^{13}$ cm$^{-2}$ and in particular
49\% of the lines have $\mbox{\novi}\leq 7.9\times 10^{12}$
cm$^{-2}$. There seems to be no correlation between the simulated
\novi and \nhi, exactly like to what is observed (right panel of the
figure), although the maximum averaged \nhi column density in the
simulations is $2.1\times 10^{19}$ cm$^{-2}$, a value a little bit
smaller than the observed one.

In the present model for $t\geq 14.5$ Myr the \ovi column densities
are smaller than $1.7\times 10^{13}$ cm$^{-2}$ and $\langle
\mbox{\novi} \rangle \leq 8.5\times 10^{12}$ cm$^{-2}$. These values
are in excellent agreement with the mean column density of
$7\times10^{12}$ cm$^{-2}$ inferred from analysis of FUSE absorption
line data in the Local ISM (Oegerle et al. 2005).

\section{Final Remarks}
The supernova-driven ISM model discussed here uses sophisticated
numerical schemes and has proven to be a valuable tool in ISM
studies. The results obtained with the ISM module in probing the
evolution of the local ISM are in excellent agreement with those
from recent observations by the FUSE satellite.

The present simulations still neglect an important component of the
ISM, i.e., high energy particles, which are known to be in rough
energy equipartition \textit{locally} with the magnetic field, the
thermal and the turbulent gas in the ISM. The presence of CRs and
magnetic fields in galactic halos is well known and documented by
many observations of synchrotron radiation generated by the electron
component. The fraction of cosmic rays that dominates their total
energy is of Galactic origin and can be generated in SN remnants via
the diffusive shock acceleration mechanism to energies up to
$10^{15}$ eV (for original papers see Krymsky et al. 1977, Axford et
al. 1977, Bell 1978, Blandford \& Ostriker 1978, for a review Drury
1983, for more recent calculations see Berezhko 1996). The
propagation of these particles generates MHD waves due to the
streaming instability (e.g.\ Kulsrud \& Pearce 1969) and thereby
enhances the turbulence in the ISM. In addition, self-excited MHD
waves will lead to a dynamical coupling between the CRs and the
outflowing fountain gas, which will enable part of it to leave the
Galaxy as a wind (Breitschwerdt et al.  1991, 1993, Dorfi \&
Breitschwerdt 2005). Furthermore, as the CRs act as a weightless
fluid, not subject to radiative cooling, they can bulge out magnetic
field lines through buoyancy forces. Such an inflation of the field
will inevitably lead to a Parker type instability, and once it
becomes nonlinear, it will break up the field into a substantial
component parallel to the flow (see Kamaya et al.\ 1996), thus
facilitating gas outflow into the halo. We are currently performing
ISM simulations including the CR and wave component.

\small{
\section*{\small Acknowledgments}

\noindent M.A. would like to thank the organization for the partial support to
attend this conference. Further support was provided by the Portuguese Science Foundation through the FAAC program under grant number FAAC/04/5/929.
}


\begin{thebibliography}{}
\bibitem[2000]{} Avillez M. A., 2000, MNRAS, 315, 479 
\bibitem[2001]{} Avillez M. A., \& Berry, D. L. 2001, MNRAS, 328, 708
\bibitem[2004]{} Avillez M. A., \& Breitschwerdt, D., 2004, A\&A, 425, 899 
\bibitem[2005]{} Avillez M. A., \& Breitschwerdt, D., 2005a, A\&A, in press [astro-ph/0502327]
\bibitem[2005]{} Avillez M. A., \& Breitschwerdt, D., 2005b, in "Astrophysics in the Far Ultraviolet - Five years of discovery with FUSE", eds. G. Sonneborn, W. Moos, \& B.-G. Andersson, ASP Conf. [astro-ph/0501466]
\bibitem[2005]{} Avillez M. A., \& Breitschwerdt, D., 2005c, ApJ, in preparation
\bibitem[2001]{} Avillez M. A., \& Mac Low, M.-M.\ 2001, ApJ, 551, L57
\bibitem[2002]{} Avillez M. A., \& Mac Low, M.-M.\ 2002, ApJ, 581, 1047
\bibitem[1977]{} Axford, W.I., Leer, E., \& Skadron, G. 1977, in Proc. 15th Int. Cosmic Ray Conf. (Plodiv) 11, 132

\bibitem[2001]{} Balsara, D. S. 2001, J. Comput. Phys., 174, 614
\bibitem[1978]{} Bell, A. R.\ 1978, MNRAS, 182, 147
\bibitem[1996]{} Berezhko, E. G. 1996, APh 5, 367
\bibitem[1989]{} Berger, M. J., \& Colella, P.\ 1989, J. Comp Phys., 82, 64
\bibitem[2002]{} Bergh\"ofer, T., \& Breitschwerdt, D.\ 2002, A\&A, 390, 299.
\bibitem[2002]{} Boldyrev, S.\ 2002, ApJ, 569, 841
\bibitem[2002]{} Boldyrev, S., Nordlund, \AA, \& Padoan, P.\ 2002, ApJ, 573, 678

\bibitem[2001]{br01} Breitschwerdt D.\ 2001, Ap\&SS 276, 163

\bibitem[1991]{} Breitschwerdt, D., McKenzie, J.F., \& V\"olk, H.J. 1991, A\&A 245, 79
\bibitem[1993]{} Breitschwerdt, D., McKenzie, J.F., \& V\"olk, H.J. 1993, A\&A 269, 54

\bibitem[2004]{} Breitschwerdt, D., \& Cox, D. P.\ 2004, in ``How does the
Galaxy Work?'', eds. E. Alfaro, E. Perez, \& J. Franco, Kluwer (Dordrecht), p. 391
\bibitem[2005]{} Cox, D. P.\ 2004, Ap\&SS, 289, 469
\bibitem[1998]{} Dai W., \& Woodward, P. R. 1998, J. Comput. Phys., 142, 331
\bibitem[2005]{} Dorfi, E.A., Breitschwerdt, D.\ 2005, A\&A, in preparation
\bibitem[1983]{} Drury, L. O'C. 1983, Rep. Prog. Phys. 46, 973
\bibitem[2003]{} Hanasz, M., \& Lesch, H.\ 2003, A\&A, 412, 331 

\bibitem[1996]{} Kamaya, H., Mineshige, S., Shibata, K., \& Matsumoto, R.\ 1996, ApJ, 458, L25 
\bibitem[1977]{} Krymsky, G. F. 1977, Dokl.~Nauk.~SSR~234, 1306, (Engl.~Trans. Sov.~Phys.~Dokl.~23, 327)
\bibitem[1969]{} Kulsrud, R. M., \& Pearce, W. D.\ 1969, ApJ, 156, 445
\bibitem[1998]{} Mac Low, M.-M., Klessen, R. S., Burkert, A., \& Smith, M. D.\ 1998, Phys. Rev. Lett., 80, 2754
\bibitem[2005]{} Oegerle, W. R., Jenkins, E. B., Shelton, R. L., Bowen, D. V., \& Chayer, P.\ 2005, ApJ, 622, 3770
\bibitem[1998]{} Porter, D. H., Woodward, P. R., \& Pouquet, A.\ 1998, Phys. Fluids, 10, 237
\bibitem[2003]{} Ryu, D., Kim, J., Hong, S. S., \& Jones, T. W.\ 2003, ApJ, 589, 338
\bibitem[1998]{} Thornton, K., Gaudlitz, M., Janka, H.-Th. \& Steinmetz, M.\ 1998, ApJ, 500, 95

\end{thebibliography}
\end{document}